\documentclass[12pt]{article}
\usepackage{rotate}
\usepackage{epsfig}
\usepackage{psfrag}
\usepackage{a4}
\begin{document}

\newcommand{\bq}{\ensuremath{{\bf q}}}
\renewcommand{\cal}{\ensuremath{\mathcal}}
\newcommand{\bqp}{\ensuremath{{\bf q'}}}
\newcommand{\bbq}{\ensuremath{{\bf Q}}} 
\newcommand{\bp}{\ensuremath{{\bf p}}}
\newcommand{\bpp}{\ensuremath{{\bf p'}}}
\newcommand{\bk}{\ensuremath{{\bf k}}}
\newcommand{\bx}{\ensuremath{{\bf x}}}
\newcommand{\bxp}{\ensuremath{{\bf x'}}}
\newcommand{\by}{\ensuremath{{\bf y}}}
\newcommand{\byp}{\ensuremath{{\bf y'}}}
\newcommand{\bxpp}{\ensuremath{{\bf x''}}}
\newcommand{\rmd}{\ensuremath{{\rm d}}}
\newcommand{\intk}{\ensuremath{{\int \frac{d^3\bk}{(2\pi)^3}}}}
\newcommand{\intq}{\ensuremath{{\int \frac{d^3\bq}{(2\pi)^3}}}}
\newcommand{\intqp}{\ensuremath{{\int \frac{d^3\bqp}{(2\pi)^3}}}}
\newcommand{\intp}{\ensuremath{{\int \frac{d^3\bp}{(2\pi)^3}}}}
\newcommand{\intpp}{\ensuremath{{\int \frac{d^3\bpp}{(2\pi)^3}}}}
\newcommand{\intx}{\ensuremath{{\int d^3\bx}}}
\newcommand{\intxp}{\ensuremath{{\int d^3\bx'}}}
\newcommand{\intxpp}{\ensuremath{{\int d^3\bx''}}}
\newcommand{\drho}{\ensuremath{{\delta\rho}}}
\newcommand{\rhoh}{\ensuremath{{\widehat{\rho}}}}
\newcommand{\fh}{\ensuremath{{\widehat{f}}}}
\newcommand{\phih}{\ensuremath{{\widehat{\phi}}}}
\newcommand{\thetah}{\ensuremath{{\widehat{\theta}}}}
\newcommand{\etah}{\ensuremath{{\widehat{\eta}}}}
\newcommand{\0}{\ensuremath{{(\bk,\omega)}}}
\newcommand{\x}{\ensuremath{{(\bx,t)}}}
\newcommand{\xp}{\ensuremath{{(\bx',t)}}}
\newcommand{\xtp}{\ensuremath{{(\bx',t')}}}
\newcommand{\xtpp}{\ensuremath{{(\bx'',t')}}}
\newcommand{\xttpp}{\ensuremath{{(\bx'',t'')}}}
\newcommand{\xtpn}{\ensuremath{{(\bx',-t')}}}
\newcommand{\xtppn}{\ensuremath{{(\bx'',-t')}}}
\newcommand{\xn}{\ensuremath{{(\bx,-t)}}}
\newcommand{\xpn}{\ensuremath{{(\bx',-t)}}}
\newcommand{\xppn}{\ensuremath{{(\bx',-t)}}}
\newcommand{\xpp}{\ensuremath{{(\bx'',t)}}}
\newcommand{\xxp}{\ensuremath{{(\bx,t;\bx',t')}}}
\newcommand{\Crr}{\ensuremath{{C_{\rho\rho}}}}

\newcommand{\Crf}{\ensuremath{{C_{\rho f}}}}
\newcommand{\Crt}{\ensuremath{{C_{\rho\theta}}}}
\newcommand{\Cff}{\ensuremath{{C_{ff}}}}
\newcommand{\Cffh}{\ensuremath{{C_{f\fh}}}}
\newcommand{\Ct}{\ensuremath{{\dot{C}}}}
\newcommand{\Ctt}{\ensuremath{{\ddot{C}}}}
\newcommand{\Crrp}{\ensuremath{{\dot{C}_{\rho\rho}}}}
\newcommand{\Crfp}{\ensuremath{{\dot{C}_{\rho f}}}}
\newcommand{\Crtp}{\ensuremath{{\dot{C}_{\rho\theta}}}}
\newcommand{\Cffp}{\ensuremath{{\dot{C}_{ff}}}}
\newcommand{\Crrpp}{\ensuremath{{\ddot{C}_{\rho\rho}}}}
\newcommand{\thetab}{\ensuremath{{\overline{\theta}}}}
\newcommand \be  {\begin{equation}}
\newcommand \bea {\begin{eqnarray} \nonumber }
\newcommand \ee  {\end{equation}}
\newcommand \eea {\end{eqnarray}}

\title{Extreme value problems in Random Matrix Theory and other disordered systems}

\author{Giulio Biroli$^{1,3}$, Jean-Philippe Bouchaud$^{2,3}$, Marc Potters$^{3}$}
\maketitle

\small{
$^1$ Service de Physique Th{\'e}orique,
Orme des Merisiers -- CEA Saclay, 91191 Gif sur Yvette Cedex, France.
\\
$^2$ Service de Physique de l'{\'E}tat Condens{\'e},
Orme des Merisiers -- CEA Saclay, 91191 Gif sur Yvette Cedex, France.
\\
$^3$ Science \& Finance, Capital Fund Management, 6 Bd
Haussmann, 75009 Paris, France.}

\begin{abstract}
We review some applications of central limit theorems and extreme values statistics in the context of disordered systems. We
discuss several problems, in particular concerning Random Matrix Theory and the generalisation of the Tracy-Widom distribution when 
the disorder has ``fat tails''. We underline the relevance of power-law tails for Directed Polymers and mean-field Spin Glasses, and we point out
various open problems and conjectures on these matters. We find that in many instances the assumption of Gaussian disorder cannot be
taken for granted.
\end{abstract}

\section{Introduction}

Most statistical models of disorder start by assume that randomness has Gaussian statistics -- from the classic Brownian motion
to the Edwards-Anderson (or Derrida) models of spin-glasses, Kraichnan models of turbulent flows, KPZ models of surface growth, 
Black-Scholes models of financial markets, etc. Thanks to the outstanding mathematical properties of Gaussian random variables, this assumption is
often technically very convenient and allows one to use powerful analytical techniques: stochastic calculus and Ito's lemma, field theory and 
replicas, etc..
The real (and often implicit) justification is however the existence of a Central Limit Theorem. This should ensure that one large enough lengths scales 
or time scales, the physical results are universal, independent of the details of the microscopic randomness -- which can therefore, for
simplicity and congeniality, be chosen as Gaussian. The paradigm of such a mechanism is the Brownian motion; in this case, provided 
elementary hops are sufficiently decorrelated from one another, it is well known that the sum of a very large number of these small
displacements leads to a Gaussian diffusion profile, quite independently of the distribution of elementary hops -- whenever its second 
moment is finite. If the second moment diverges, the walk becomes a L\'evy flight, with anomalous diffusion described 
by the generalized Central Limit Theorem of L\'evy and Gnedenko which again ensures a certain degree of universality \cite{PhysRep}: only the extreme tails of the microscopic distribution matter 
in the macroscopic limit. Although this dichotomy between finite and infinite variance is asymptotically rigorous, finite time or size
effects can be strong and lead to effective violations of these Central Limit Theorems. An important example is when the distribution
of elementary hops has a finite variance but power-law tails. In this case, fat tail effects are persistent and 
convergence towards Gaussian diffusion is very slow. This is particularly relevant in finance, where significant deviations from
Gaussian statistics are observed even for long time lags \cite{book}.

Sums of $N$ {\sc iid} random variables therefore provide a beautiful illustration of universality and universality classes, a 
concept that extends far beyond this simple, exactly soluble example. The existence of generalized 
Central Limit Theorems for more complicated (non linear) problems involving random variables should be generic, again
leading to some universality -- universality classes should however be determined on a case by case basis and might be different
from the L\'evy-Gnedenko classification. A well known example is the statistics of extreme values, say the largest of $N$ 
independent random variables $x_i$. In this case again, the limiting distribution becomes to some degree universal; one has to 
distinguish three different cases, depending on the `microscopic' distribution $p(x)$: Weibull (for distributions $p(x)$ 
which strictly vanish beyond a finite value $x^*$), Gumbel-Fisher-Tippett (for distributions decaying faster than any power-law) and Fr\'echet
(for power-law distributions)\cite{Galambos}. Interestingly, it is possible to formulate a problem which interpolates between {\it sums} of random 
variables and {\it extremes} of random variables, by considering the following quantity:
\be
S_q = \left[ \sum_{i=1}^N x_i^q \right],
\ee
where one assumes for simplicity that $x_i$'s are all positive. Clearly, $q=1$ corresponds to a simple sum, whereas when $q \to \infty$
at fixed $N$, $S_q^{1/q}$ converges to the largest element $x_{\max}$. Defining $x_i \equiv \exp-\varepsilon_i$, it is clear that $S_q$ 
plays the role of a partition function and $q$ is the inverse temperature for a generalized Random Energy Model (REM), where the energies 
$\varepsilon_i$'s are not necessarily Gaussian. This problem was considered in \cite{BM,GBA} and has, beyond the REM interpretation, many 
different applications. For example, suppose $\varepsilon_i$ is a growth rate of specie $i$ in the population, 
or the return of asset $i$ in a portfolio, and $q$ is the time. Then $S_q$ is the total population after time $q$ or the total value 
of the portfolio after time $q$; the detailed statistics of these objects is therefore quite interesting.\footnote{
More complex situations,  
for example diffusion of species in a random environment, can be analyzed along similar lines \cite{GBA2}.}The result depends on the 
relative value of $q$ and $N$ when both diverge to infinity. More precisely, taking for simplicity $\varepsilon_i$ to be Gaussian with
variance $\sigma^2$, the relevant parameter is $\mu=\sqrt{2\ln N}/q \sigma$. The statistics of $S_q(N)$ only depends on $\mu$ and, 
quite interestingly, closely follows the above Gauss/L\'evy dichotomy: for $\mu >2$, $S_q$ is Gaussian; for $\mu < 2$ it becomes
L\'evy distributed and more and more dominated by extreme values. In fact, as soon as $\mu < 1$, the whole sum $S_q$ is well approximated
by a finite number of terms, whereas when $\mu \to 0$, only the largest survives. The transition at $\mu =1$ corresponds
exactly the glass transition in the REM. The above results can be extended to any distributions of $\varepsilon_i$ in the Gumbel class, 
up to a redefinition of $\mu$ \cite{BM,GBA}; interestingly, the detailed statistics of $S_q$ is precisely encoded in the 1-step Replica Symmetry 
Broken solution of the generalized Random Energy Model \cite{Derrida,BM}.

As the above example illustrates, it is clear that low temperature/long time properties of disordered systems are sensitive to extreme
values rather than to typical values; this change of focus means that one should {\it a priori} be specially weary about universality
classes and the influence of the choice of distribution on the physical results. The aim of this paper is to discuss several problems
within this perspective, reviewing some recent and older results, and pointing out several open problems and technical challenges well worth investigating 
in the future.

\section{Random Matrices and Top Eigenvalues}

A remarkable example of universal limit distribution is the eigenvalue spectrum of random symmetric $N \times N$ matrices ${\bf M}$ 
with {\sc iid} real elements. Again, as soon as the variance of the matrix elements is finite, the eigenvalue density $\rho(\lambda)$ 
converges to the Wigner semi-circle, with edges at $\lambda= \pm 2$ when the variance of the entries is normalized to $1/N$. This 
result can be derived in a way which makes direct use of the Central Limit Theorem for sums of random variables, and in this way
makes explicit the mechanism underpinning the universality of the Wigner semi-circle. In line with the above classification, one 
finds that where the variance of entries diverge, the eigenvalue spectrum $\rho(\lambda)$ of ${\bf M}$ is no 
longer the Wigner semi-circle. Not surprisingly, the discussion parallels the L\'evy-Gnedenko classification for sums of random 
variables.  The case where the distribution of entries decays as a power-law $\sim |M_{ij}|^{-1-\mu}$ 
(possibly multiplying a slow function) with $\mu < 2$ (such that the variance of entries diverge), the eigenvalue spectrum 
$\rho(\lambda)$ can be computed exactly \cite{CB,Burda} and has no longer a compact support, but itself acquires power-law tail 
$\rho(\lambda) \sim |\lambda|^{-1-\mu}$, bequeathed from the tails of the matrix entries \cite{CB}. The structure of eigenvectors
is also quite interesting: whereas for $\mu >2$ most states are extended, various localisation transitions occur, as a function of
$\lambda$, for $\mu < 2$.

Following the above discussion, it is quite natural to investigate the distribution of extreme eigenvalues as well, and
to find the universality classes corresponding to this question. Since the eigenvalues of a random matrix are strongly correlated
random variables, one does not expect the result to belong to any of the well known Gumbel-Fisher-Tippett, Weibull and
Fr\'echet cases. One of the most exciting recent result in mathematical physics is the Tracy-Widom distribution of the 
top eigenvalue of large Gaussian random matrices \cite{TW}. The truly amazing circumstance is that the very same distribution 
also appears in a host of physically important problems \cite{Spohn}: crystal shapes, exclusion processes \cite{Satya}, 
sequence matching, and, as discussed in the next section, directed polymers in random media.

Again, the Tracy-Widom result is expected to hold for a broad class of random matrices. The precise 
characterisation of this class, as well as the extension of the Tracy-Widom result for other classes, is a subject of 
intense activity \cite{Sosh1,BBAP}. The case where the distribution of entries decays as a power-law 
$\sim |M_{ij}|^{-1-\mu}$ is expected to fall in a different universality class, at least when $\mu$ is small enough. 
The situation is simple when $\mu <2$: for large $\lambda$, eigenvalues become uncorrelated and, as mentionned above, 
distributed as $\rho(\lambda) \sim |\lambda|^{-1-\mu}$. Correspondingly, the largest 
eigenvalues are described by Fr\'echet statistics \cite{Sosh2}. What happens when $\mu$ is in the range $]2, +\infty)$, such that 
the eigenvalue spectrum $\rho(\lambda)$ still converges \cite{CB}, for large $N$, to the Wigner semi-circle? We find that
Fr\'echet statistics holds whenever $\mu < 4$, whereas the Tracy-Widom applies asymptotically as soon as $\mu > 4$, with a
new limiting family of distributions for $\mu=4$ \cite{BBP}. The idea of the method is to start with a real symmetric matrix $\widehat{\bf M}$ 
with {\sc iid} elements of variance equal to $1/N$, and such that the distribution has a tail decaying as: 
\be\label{dist}
p(M_{ij}) \simeq \frac{\mu (A N^{-1/2})^\mu}{|M_{ij}|^{1+\mu}},
\ee
where the tail amplitude ensures that $M_{ij}$'s are of order $A N^{-1/2}$. As soon as $\mu > 2$, the density of eigenvalues converges 
to the Wigner semi-circle on the interval $\lambda \in [-2,2]$, meaning that the probability to find an eigenvalue 
beyond $2$ goes to zero when $N \to \infty$. However, this does not necessarily mean that the largest eigenvalue 
tends to $2$ -- we will see below that this is only true when $\mu > 4$.  Now, we perturb the matrix $\widehat{\bf M}$ 
by adding a certain amount $S$ to a given pair of matrix elements, say $(\alpha,\beta)$: 
$\widehat M_{\alpha \beta} \to \widehat M_{\alpha \beta} + S$ and $\widehat M_{\beta \alpha} \to \widehat M_{\beta \alpha} + S$. 
What can one say about the spectrum of this new matrix? Using self-consistent perturbation theory, which becomes exact for 
large $N$, one can show \cite{BBP} that when $|S| \ge 1$, there is a pair of eigenvectors partly localised on $\alpha,\beta$, with
eigenvalues $\lambda_\pm=\pm(S+1/S)$ with $|\lambda| \geq 2$, which is therefore expelled from the Wigner sea.  When $|S| <1$, one 
the other hand, no such eigenvalue exist and the edge of the spectrum remains $\lambda_{{\max}}=2$ in this case. 

\begin{figure}
\begin{center}
\psfig{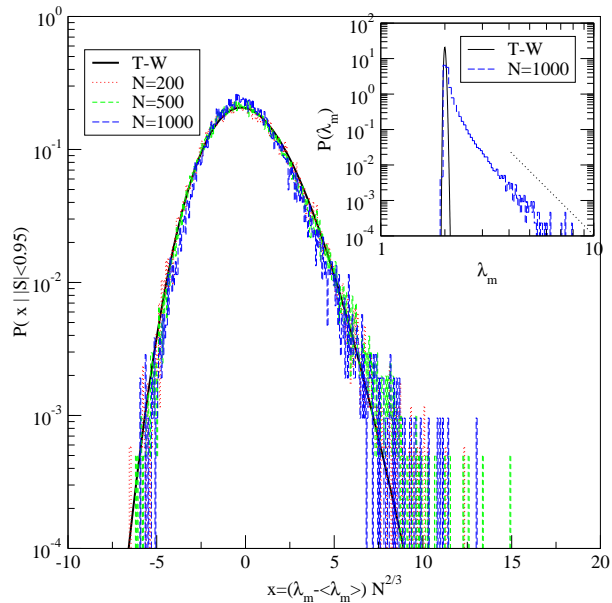} 
\end{center}
\caption{Histogram of $\lambda_{{\max}}$ conditioned on
$|S|<0.95$ for $\mu=5$ for $N=200,500,1000$ each eigenvalue has been
shifted by the empirical mean and scaled by $N^{2/3}$, for comparison
a GOE Tracy-Widom distribution of zero mean and variance adjusted to
match $N=500$ data is also shown (data obtained from
\cite{Prahofer}). Similar agreement with Tracy-Widom and scaling in
$N^{2/3}$ is obtained for any value of $\mu$ when conditioned on
$|S|<0.95$. Note that for the parameters chosen here, the probability of
$|S|>1$ is still quite large (75.2\%) at $N=1000$. Even for such
large values of $N$, the unconditional distribution of
$\lambda_{{\max}}$ has a marked power-law tail of index $\mu$ (dotted line) and is very
different from the asymptotic Tracy-Widom distribution
(inset).}
\end{figure}

Now consider large matrix entries $|M_{ij}| > 1$. From Eq. (\ref{dist}),
their number is $N^2 \int_1^\infty p(M_{ij}) dM_{ij} = A^\mu
N^{2-\mu/2}$. In the case $\mu > 4$, it is clear that this number
tends to zero when $N \to \infty$. With probability close to unity for large $N$, no
entry is larger than one, in which case the largest eigenvalue is expected to remain Tracy-Widom
around $\lambda^*=2$. With small probability, however, the
largest element $S$ of ${\bf M}$ exceeds one; its distribution is
$A^\mu N^{2-\mu/2}/|S|^{1+\mu}$ and the corresponding largest
eigenvalue, using the above analysis, is $\lambda_{{\max}}=S+1/S$. For
$\mu > 4$ and large but finite $N$, we therefore expect that the
distribution of the largest eigenvalue of ${\bf M}$ is Tracy-Widom,
but with a power-law tail of index $\mu$ that very slowly disappears
when $N \to \infty$. Our numerical results are in full agreement with
this expectation (see Fig. 1). When $\mu < 4$, on the other hand, the
number of large entries increases with $N$. However, when $\mu$ is
larger than $2$, such as to ensure that the eigenvalue spectrum still
converges to the Wigner semi-circle, the number of row or columns
where two such large entries appear still tends to zero, as
$N^{2-\mu}$. Therefore, the above analysis still holds: for each
large element $S_{ij}$ exceeding unity, one eigenvalue
$\lambda=S_{ij}+S_{ij}^{-1}$ will pop out of the Wigner sea.  Even if
the eigenvalue {\it density} tends to zero outside of the interval
$[-2,2]$ when $2 < \mu < 4$, the {\it number} of eigenvalues exceeding
$2$ (in absolute value) grows as $N^{2-\mu/2} \ll N$.  The largest eigenvalues are then equal 
to the largest entries and are themselves given by a Poisson point process with
Fr\'echet intensity, as proven by Soshnikov in the case $\mu < 2$
\cite{Sosh2}.  His result therefore holds in the whole range $\mu <
4$. Finally, the marginal case $\mu=4$ is easy to understand from the
above discussion. The number of entries exceeding one remains of order
unity as $N \to \infty$; the distribution of the largest entry $S$ is
Fr\'echet with $N$-independent parameters: \be P_{\mu=4}(|S|)=\frac{4
A^4}{|S|^{5}} \exp\left[-\frac{A^4}{|S|^{4}}\right].  \ee

The probability that $|S|$ exceeds $1$ is $\varphi=1-e^{-A^4}$, in which case $\lambda_{{\max}}=|S|+|S|^{-1}$; 
otherwise, with probability $1-\varphi$, $\lambda_{{\max}}=2$. This characterizes entirely the asymptotic distribution 
of the largest eigenvalue in the marginal case $\mu=4$: it is a mixture of a $\delta$-peak at $2$ and a
transformed Fr\'echet distribution. Note that this asymptotic distribution is non-universal since it depends 
explicitly on the tail amplitude $A$. Again, all these results are convincingly borne out by numerical simulations, 
see \cite{BBP}. The statistics of the second, third, etc. eigenvalues could be understood along the same lines.

One can also consider the case of sample covariance matrices, important in many different contexts. The `benchmark' spectrum of
sample covariance matrix for {\sc iid} Gaussian random variables is well known, and given by the Mar\v{c}enko-Pastur distribution
\cite{MP}. Here again, the spectrum has a well defined upper edge, and the distribution of the largest eigenvalue is 
Tracy-Widom (see e.g. \cite{BBAP}). What happens if the random variables have heavy tails? More precisely, we consider 
$N$ times series of length $T$ each, denoted $x_i^t$, where $i=1,..,N$ and $t=1,...,T$. The $x_i^t$ have zero mean and 
unit variance, but may have power-law tails with exponent $\mu$. For example, daily stock returns are believed to have
heavy tails with an exponent $\mu$ in the range $3-5$ \cite{book}. The empirical covariance matrix $\bf C$ is defined as:
\be
C_{ij} = \frac{1}{T} \sum_t x_i^t x_j^t.
\ee
When the time series are independent, and for $T$ and $N$ both diverging with a fixed ratio $Q=T/N$, 
the eigenvalues of $\bf C$ are distributed in the interval $[(1-Q^{-1/2})^2,(1+Q^{-1/2})^2]$. 
When $T \to \infty$ at fixed $N$, all eigenvalues tend to unity, as they should since the 
empirical covariance matrix converges to its theoretical value, the identity matrix. When $N$ and $T$ 
are large but finite, the largest eigenvalue of $\bf C$ is, for Gaussian returns, a distance $\sim N^{-2/3}$ away from
the Mar\v{c}enko-Pastur edge, with Tracy-Widom fluctuations. When returns are accidentally large, this may cause 
spurious apparent correlations and substantial overestimation of the largest eigenvalue of $\bf C$. Let us
be more specific and assume, as above, that one particular return, say $x_\alpha^\tau$, is exceptionaly large, 
equal to $S$. A generalisation of the above self-consistent perturbation theory shows that
whenever $S \leq (NT)^{1/4}$, the largest eigenvalue remains stuck at $\lambda_{{\max}}=(1+Q^{-1/2})^2$, whereas
when $S > (NT)^{1/4}$, the largest eigenvalue becomes:
\be
\lambda_{{\max}}=\left(\frac{1}{Q}+\frac{S^2}{T}\right)\left(1+\frac{T}{S^2}\right);
\ee
This result again enables us to understand the statistics of $\lambda_{{\max}}$ as a function of the tail exponent $\mu$.
For $N$ times series of {\sc iid} random variables, of length $T$ each, the largest element is of order $(NT)^{1/\mu}$.
For $\mu > 4$, this is much smaller than $(NT)^{1/4}$ and, exactly as above, we expect the largest eigenvalue of $\bf C$
to be Tracy-Widom, with possibly large finite size corrections \cite{Burda2}. For $\mu < 4$, large `spikes' in the time series 
dominate the top eigenvalues, which are of order $\lambda_{{\max}} \sim N^{4/\mu-1} Q^{2/\mu-1}$ and distributed 
according to a Fr\'echet distribution of index $\mu/2$.  In the marginal case 
$\mu=4$, as above, $\lambda_{{\max}}$ has a finite probability to be equal to the Mar\v{c}enko-Pastur value, and with 
the complementary probability it is distributed according to a transformed Fr\'echet distribution of index $2$, with a $T$ and 
$N$ independent scale. The structure of the corresponding eigenvectors can also be investigated and is again found to be partly 
localized when $S > (NT)^{1/4}$. Finally, we expect similar results to hold for the Random Singular Value problem studied in 
\cite{Augusta}, where rectangular matrices corresponding to cross correlations between different sets of time series are considered. 

\section{Directed Polymers, KPZ/KPP Equations and Pinned Manifolds} 

Quite remarkably, the Tracy-Widom distribution for the largest eigenvalue of complex sample covariance 
matrices has deep links with the Directed Polymer (DP) problem in (1+1) dimension, defined as follows: consider a two 
dimensional square lattice, such that on each site one independently draws a random energy $e(x,y)$ from a given distribution $p(e)$.
The zero-temperature directed polymer is the directed walk starting from $(0,0)$ and only allowed to move North-East, such that the
sum of encountered energies is minimum. When $e$ is an exponential variable, there is an exact mapping to the Tracy-Widom 
problem \cite{Johansson}. Thanks to this mapping, the DP in (1+1) dimension can be considered as a rare example of exactly soluble disordered system in finite 
dimensions, for which not only the scaling exponents but the full distribution of the ground state energy can be completely 
characterized in terms of the Tracy-Widom distribution. 
In particular, the scaling between the sample-to-sample fluctuations of the ground state energy $\Delta E$ and the length of the 
path $L$ is $\Delta E \sim L^{1/3}$ and the typical width of the optimal path is given by $W \sim L^{2/3}$. These conjectured scalings,
based on physical arguments, are therefore exact for a certain class of random energy distributions $p(e)$. In the spirit of the above 
discussion, it is quite natural to wonder about the universality class of such results \cite{Zhang}. 

The question is all the more interesting that the DP problem maps onto a non-linear stochastic partial differential equation 
describing non-equilibrium surface growth, the so-called Kardar-Parisi-Zhang equation \cite{HH}:
\be
\frac{\partial h(\vec r,t)}{\partial t} = \nu \Delta h + \frac{\Lambda}{2} (\vec \nabla h)^2 + \eta(\vec r,t),
\ee
where $h$ is the height of the interface, $\vec r$ the coordinates along the $d$-dimensional interface and $\eta$ a white noise
term, mapping to the random energy $e$ in the DP problem. (The (1+1) DP problem corresponds to $d=1$). This equation can in turn 
be interpreted as a Burgers equation on the quantity $\vec u = -\vec \nabla h$, which corresponds to the effective 
force acting on the end point of the directed polymer. But since the Burgers equation is a toy-model for turbulence, the question
of the relevance of the distribution of the external forcing term $\vec \nabla \eta$ on the statistics of the velocity field is particularly
interesting. KPZ-like equations may also describe very different physical situations, such as, for example, propagation of crack 
fronts in disordered materials \cite{BBLP}; the robustness of the results with respect to the nature of the randomness is therefore quite important. 

It is clear that it is the presence of a non-linear term in the KPZ/Burgers equation, responsible for the 
appearance of shocks in the velocity field, which makes the problem non trivial. For a linear KPZ equation $\Lambda=0$, called
the Edwards-Wilkinson equation in this context, it is
easy to show that the long time statistics of $h(\vec r,t)$ is Gaussian provided $\eta(\vec r,t)$ has a finite second moment, 
again thanks to the Central Limit Theorem for sums. The interplay between non-linearity and fat tails does however lead to 
rather unexpected results, as suggested by a simple Flory argument in the case where $p(e)$ decays, for large negative $e$, as
$p(e) \sim |e|^{-1-\mu}$. The Flory argument compares the energy of the best `bounty' site in a volume 
$V = W^d L$ to the elastic stretching energy the polymer has to pay to get there, of order $W^2/L$. Using extreme value theory in the
Fr\'echet case, one gets $E_{\min} \sim - V^{1/\mu}$, leading to $W \sim L^{(1+\mu)/(2\mu-d)}$ and $\Delta E \sim E_{\min} \sim L^{(2+d)
/(2\mu-d)}$.
Of course, this reasonning is only valid if the extreme bounty site is worth the trip, i.e., if the distortion $W$ is larger than 
in the absence of tails in the distribution of $e$. Calling $\zeta_d$ the exponent relating $W$ to $L$ in the `thin tail' case, one
expects the results to be strongly affected by the power-law tail of $p(e)$ as soon as $\mu$ is smaller than:
\be\label{muc}
\mu < \mu_c = \frac{1+d\zeta_d}{2\zeta_d-1}.
\ee
The exponent $\zeta_d$ is only known exactly in $d=1$, with only numerical estimates available in $d >1$ and still a very controversial
situation concerning the value of the upper critical dimension $d_c$ above which $\zeta_d$ takes the trivial random walk 
value $\zeta=1/2$. In any case, for $d=1$, $\zeta_1=2/3$, leading to $\mu_c = 5$. In other words, the Flory argument suggests that
as soon as the fifth moment of the local energy distribution diverges, the Tracy-Widom scaling breaks down, contrarily to the 
naive expectation that the DP exponents are universal as long as the variance of the local disorder is finite. The value $\mu_c=5$
is also different from the critical value $\mu_c=4$ found for the statistics of the top eigenvalue of random matrices, showing that
a general mapping between the two problems, if it exists, is more subtle.

Summarising, the Flory argument suggests that in (1+1) dimensions, the energy fluctuations should scale as $L^{1/3}$ and by Tracy-Widom 
for $\mu > 5$, and as $L^{3/(2\mu-1)}$ for $2 < \mu < 5$ with a new type of limiting distribution 
(the case $\mu < 2$ corresponds to a complete stretching of the polymer and was recently solved in \cite{Hamley}). We have 
conducted new numerical simulations of this problem which indeed confirm that for $\mu > 5$, the ground state energy 
scales as $L^{1/3}$ with Tracy-Widom fluctuations, while for $\mu < 5$ the above Flory prediction is very accurate. The 
distribution $P$ of ground state energy can be well fitted by a geometric convolution of Fr\'echet distributions, suggested by
the Flory argument with a finite number of dominant sites: $P=(1-p)(F+p F \star F+p^2 F \star F \star F +...)$, different from the pure Fr\'echet distributed reported above for 
the largest eigenvalue for $\mu < 4$. 

At this stage, a number of comments should be made. First, Eq. (\ref{muc}) shows that $\mu_c$ diverges when $d \to \infty$. This is
perfectly in line with the Derrida-Spohn solution for the DP on a tree \cite{DS}, which indeed breaks down completely as soon as $p(e)$ decays 
slower than exponentially when $e \to \infty$. This, in turn, is related to the problem of fronts in the Kolmogorov-Petrovsky-Piscounov (KPP) 
equation when the initial condition decays too slowly into the unstable phase. Whereas for localised initial conditions, the front between the
stable and unstable phase propagates at a well defined velocity (which determines the free-energy of the DP \cite{DS,PLDlog}), the case of slowly
decaying initial conditions has, to our knowledge, not been carefully investigated. The very notion of a propagating `front' might
disappear altogether, much like in models of epidemic propagation with infected individuals performing L\'evy flights, discussed
in this context in \cite{Brock}. 

Another intriguing property of Eq. (\ref{muc}) is that $\mu_c$ diverges whenever $\zeta_d=1/2$, independently of dimension. If a 
critical dimension exists such that the strong disorder fixed point has $\zeta_d=1/2$ for $d >d_c$, then this fixed point 
should be unstable against power-law tailed disorder with arbitrarily high exponent $\mu$. Since all moments of the disorder 
still exist in this case, it is difficult to see how perturbative methods can grasp such a strange behaviour. More generally, it is not 
obvious to see how perturbative renormalisation group methods for pinned systems (including the Functional RG) can deal with 
high moment anomalies which change the scaling exponents. This seems to us to be a very interesting technical challenge; its
resolution might indirectly shed light on the still elusive nature of the strong disorder fixed point of the DP/KPZ problem in 
high dimensions \cite{Canet}. A case where progress is possible is the problem of pinned manifold in the mean-field limit, which maps onto 
a deterministic Burgers equation for the effective force acting on the manifold \cite{BM}. Disorder is now entirely contained in the 
initial condition, which represents the (bare) microscopic pinning force. The statistics of the renormalized force, in particular
the density and amplitude of the `jumps' (Burgers shocks) responsible for the famous cusp in the renormalized correlation function 
predicted by the Functional RG, follows in this case the simple extreme value classification for independent random variables \cite{BM,Kida}. 
In particular, the case of a Gaussian pinning field is indeed in a different universality class than any power-law tailed disorder.
Interestingly, the Functional RG has recently been casted in the form of a functional Burgers equation for the effective force 
in full generality \cite{PLD}; this might provide a way to understand the importance of the statistics of the microscopic disorder in the
general case as well.

Finally, the presence of so-called `temperature chaos' \cite{FH,YS} might be quite sensitive to power-law tails in the microscopic disorder.
Temperature chaos is associated with the fact that the free energy of the DP at non zero temperatures scales as $L^{1/3}$, whereas
both energy and entropy fluctuations are dominated by small scale fluctuations and therefore scale as $L^{1/2} \gg L^{1/3}$. This implies that
when temperature changes by a small amount $\delta T$, polymers of length $L >L^*$, with  $\delta T L^{*1/2} = L^{*1/3}$ must 
rearrange and find another equilibrium configuration. One finds $L^* \sim (\delta T)^{-1/6}$, in good agreement with numerical 
work \cite{YS} and exact results on Migdal-Kadanoff lattices \cite{RdS}. Interestingly, as long as $\mu > 2$, entropy fluctuations should still scale
as $L^{1/2}$, whereas free-energy scale as  $L^{3/(2\mu-1)}$ when $\mu < 5$. Hence, when $\mu < 7/2$, small temperature changes, or
small disorder changes at zero temperature, are not strong enough perturbations to induce large scale rearrangements.

\section{Diffusion in a random potential}

On the failure of perturbative RG to grasp the potential relevance of ``fat tails'', it is interesting to mention the case of 
Langevin diffusion in a random potential, i.e., the long-time behaviour of $\langle x^2(t) \rangle$, where $\vec x$ obeys:
\be
\frac{d\vec x}{dt}= -\vec \nabla U(\vec x) + \vec \eta(\vec x,t);
\ee
where $\eta$ is again a thermal noise and $U(\vec x)$ a random potential with short range correlations, over a length $\xi$. When $U$ is Gaussian, one 
can show that diffusion is always normal, i.e.  $\lim_{t \to \infty} \langle x^2(t) \rangle/t = D > 0$. 
Exact formulas for the value of $D$ are available in 1 and 2 dimensions. Perhaps surprisingly, these formulas are exactly reproduced by a simple RG scheme, as independently proposed by
Deem and Chandler \cite{Deem}, and Dean, Drummond and Horgan \cite{Dean1}. Physically, the time spent in a given region of space
is $\tau(\vec x)\propto \exp U(\vec x)/T$; because the potential has Gaussian tails the average of $\tau(\vec x)$ is finite, and hence
the diffusion normal, with $D \sim \xi^2/\langle \tau \rangle$. But as soon as the disorder has tails decaying slower than exponential, the same argument leads to an infinite
average trapping time and to subdiffusion. In the special case where $p(U) \sim \exp(-U/T_g)$, the model is similar to the trap model
\cite{MB,GBAC}; one expects to find a critical temperature below which diffusion becomes anomalous, and $\langle x^2(t) \rangle \sim t^\mu$
with $\mu < 1$. This result was indeed recently confirmed exactly \cite{Dean2}; however, the RG scheme which leads to such an accurate result for 
Gaussian potentials is totally blind to the tails of the disorder, and is unable to reproduce the above phase transition and 
subdiffusive behaviour. It would be very interesting to see how to adapt the RG scheme of \cite{Deem,Dean1} to subexponential tails; 
this might be an important technical breakthrough to address quantitatively the issue of activated dynamics in supercooled glasses
below the Mode-Coupling temperature.

\section{Spin Glasses and non-standard RSB}

In this final section we address the emblematic spin-glass problem, in particular the mean-field Sherrington-Kirkpatrick (SK) model,
from the point of view of extreme value statistics. The SK model is defined by the following mean-field Hamiltonian:
\be
{\cal H} = - \frac{1}{\sqrt{N}} \sum_{i,j} J_{ij} S_i S_j, \qquad S_i = \pm 1,
\ee
where $J_{ij}$ are independent Gaussian random variables. It is now proven mathematically that the full Replica Symmetry Broken 
solution invented by Parisi provides the exact solution fo the SK model and encodes in a rather magical way the complexity of the spin-glass
phase \cite{MPV}. Much insight into the meaning of Parisi's hierarchical construction is provided both by the cavity approach and by the
Random Energy Model. One understands that RSB is essentially a algebraic way to capture the Gumbel-Fisher-Tippett statistics of 
low lying energy states, which become relevant below the spin-glass transition \cite{BM}. The symptoms signalling that the replica symmetric (RS), one phase 
solution is unstable are well known: its entropy becomes negative below a certain temperature (and remains negative at zero temperature), 
and the spin-glass susceptibility, which measures the correlation of spin fluctuations, assumed to be negligible in the RS description,
in fact diverges below the spin-glass transition temperature. One of the striking prediction of the RSB solution is the pseudo-gap in the distribution
of local magnetic fields $h$, which vanishes linearly when $h \to 0$: $P(h) \sim A h$ at $T=0$, in contrast with the Gaussian 
distribution of $h$ found in the RS phase. Interestingly, this pseudo-gap is intimately related to the marginal stability of the
ground state configuration \cite{Anderson}, and is the close analog of the Efros-Shklovskii gap in Coulomb glasses \cite{Muller}. As a consequence of the vanishing
of $P(h)$ for small $h$'s, the specific heat of the spin-glass grows as $T^2$ at small $T$, rather than the naively expected linear-in-$T$
behaviour. 

How are these results affected by non Gaussian disorder? It is easy to be convinced, for example using the cavity method \cite{Parisi}, that the 
SK results are universal provided the central limit theorem holds, i.e., the variance of the $J_{ij}$s is finite and correlations
can be neglected. One knows that introducing strong correlations between the $J_{ij}$s, as is the case of the Random Orthogonal 
Model, may change the nature of the solution, for example from full RSB to one-step RSB \cite{ROM}. When the $J_{ij}$s are {\sc iid} variables
with a power law tail $|J_{ij}|^{-1-\mu}$ with $\mu < 2$ such that the variance is infinite, interesting effects appear, related to
the fact that some bonds become extremely strong compared to others, thereby decreasing the frustration in the system. For 
example, there is a `trivial' spin glass phase, i.e. a phase where the Edwards-Anderson parameter is non zero and the RS solution is
stable \cite{LSG}. RS is broken below a second transition temperature $T_{AT}$; however the precise nature of the RSB phase is still unknown. 
The puzzle is that one naively expects the distribution of low-energy states to be governed by Fr\'echet statistics, since the energy is a 
sum of power-law distributed random variables. One knows from the study of the REM with power-law distributed energies that the
solution in that case cannot be the standard Parisi replica symmetry breaking scheme \cite{BM}, which -- as mentionned above -- describes the 
Gumbel universality class.\footnote{The same remark applies to problems where the ground state energy is strictly bounded, such as
the number partionning problem, falling into the Weibull universality class \cite{Mertens}.} Things might however be more subtle: the energy of low lying states might be of the form $E_0 + e_\alpha$,
where $E_0$ is indeed Fr\'echet but common to all states, and a Gumbel residue $e_\alpha$, restoring the Parisi scheme. This should 
be associated with a distribution of overlaps $q$ which does not extend down to $q = 0$.

Even in the absence of a detailed solution below $T_{AT}$, one can argue \cite{LSG} that the distribution of local fields again develops a gap when $1 < \mu 
< 2$, of the form  $P(h) \sim A_\mu h^{\mu-1}$. Correspondingly, the specific heat now grows as $T^\mu$. The situation becomes 
quite interesting below $\mu = 1$ since in this case replica symmetry appears to be restored at zero temperature, and the gap in
$P(h)$ disappears. The hierarchy between bonds is so strong for $\mu < 1$ that only very weak bonds appear to be frustrated in 
that case, unable to generate a large number of different states. It would be very interesting to characterize in detail the solution
of this `L\'evy' SK model, and decide whether or not one needs to invent a non-standard RSB scheme for this problem.

We again end this section by multifarious remarks. First, the Tracy-Widom problem of the largest eigenvalue of Random Matrices can
be seen as the statistics of the ground state energy of the $p=2$ spherical spin-glass. It is very natural to ask about the 
statistics of the ground state energy of the (hard-spin) SK model. Interestingly, this problem is still open. Numerically, it is
known that the {\it average} ground state energy per spin converges towards the asymptotic, $N = \infty$ result predicted by the Parisi solution
with $N^{-2/3}$ corrections, as for the Tracy-Widom problem. Whether or not this is a coincidence is not understood; one should note 
that the sample-to-sample fluctuations of this ground state energy are {\it not} of order $N^{-2/3}$ (and not described by a Tracy-Widom
distribution \cite{ABM}) but of order $N^{-\omega}$ with $\omega \approx 3/4 > 2/3$. Again, there are no theoretical consensus on the value of $\omega$ 
\cite{BKM,Asp}; only qualitative arguments for $\omega \equiv 3/4$ exist. The generalisation of these arguments to 
L\'evy spin-glasses suggest $\omega = 1 - 1/\mu^2$; it would be interesting to have numerical data to test this prediction. 
Another natural extension of the Tracy-Widom problem within the context of spin-glasses is the statistics of the ground state 
energy of the spherical $p$-spin glass, which leads to a non-linear eigenvalue problem. In the $p=3$ case, one should find the 
largest eigenvalue of:
\be
\sum_{jk=1}^N J_{ijk} \phi_j \phi_k = E \phi_i,
\ee
where $J_{ijk}$ are Gaussian random variables of variance $1/N^2$. It would be interesting to compute the scaling exponent describing
the fluctuation of this new type of extreme value problem. Could this problem also be mapped to the directed polymer problem in higher 
dimension?

Finally, we note that in finite dimensions, the role of the distribution of $J_{ij}$ on the universality class of the spin-glass transition was numerically 
investigated by Campbell et al. \cite{Campbell}; the consensus is however that (at least for fast-decaying) distribution, the critical exponents
should be universal, with possibly large sub-leading corrections. However, the situation might change for power-law distributed 
couplings, as suggested by a Migdal-Kadanoff approach to the excitation energy exponent \cite{BKM}.

\section{Conclusion and open problems}

In summary, we reviewed several examples where the detailed shape of the distribution of randomness matters more than naively
anticipated on the basis of the central limit theorem. The tail of the distribution is obviously important when one is concerned 
by extreme value problems, such as the statistics of the largest eigenvalue of heavy tailed random matrices, a problem we 
have discussed in detail. Another problem where the tails of the noise are crucial is the problem of barrier crossing and the
Arrhenius law. This is not surprising: since barrier crossing is itself a rare event, its occurence may be much enhanced by the
present of anomalous, non-Gaussian tails in the thermal noise. Note that the relevance of extreme value statistics to barrier height in disordered
systems dates back to Rammal \cite{Rammal}, see also \cite{Vinokur}. 

More surprising, at least at first sight, is the importance of these tails for the statistics of fields 
obeying a non-linear differential equation, such as the Burgers, KPZ or KPP equations, either with a stochastic forcing or with 
a stochastic initial condition. This includes the example of the directed polymer in random environment or diffusion in a random potential; 
the scaling exponents of these problems is extremely sensitive to the tails of the disorder, an unusual feature from a field
theoretical point of view. We have insisted on the technical challenges associated with this phenomenon: how should one generalize the (Functional) RG to account for non-Gaussian tails? 
Since the problem of pinned manifold can also be addressed using a replica field theory, one can similarly wonder if and how the Parisi
replica symmetry breaking scheme has to be generalised in these cases. We have underlined several other open problems and conjectures, 
such as the general relation between top eigenvalues and directed polymers, the statistics of the ground state energy of the SK model, 
the nature of the low temperature phase of L\'evy spin-glasses, etc. The solution to these problems, beyond their mere technical 
interest, could shed some light on the subtleties and surprises of the physics of disordered systems.

\vskip 1cm
Acknowledgements: We thank G. Ben Arous, S. Boettcher, Z. Burda, D. Dean, A. Lefevre, S. Majumdar, O. Martin \& M. M\'ezard for many useful 
discussions on these matters.


\begin{thebibliography}{99}

\bibitem{PhysRep} see e.g. J.-P. Bouchaud, A. Georges, {\it Anomalous diffusion in random media: statistical mechanisms,
models and physical applications}, Phys. Rep. {\bf 195} 127 (1990) and {\it L\'evy Flights and Related Topics in Physics}, 
Edts: M. F. Shlesinger, G. M. Zaslavsky, U. Frisch, Lecture Notes in Physics, vol. 450.
\bibitem{book} see e.g J.-P. Bouchaud and M. Potters, {\it Theory of Financial Risk and Derivative Pricing}, Cambridge University
Press (2004)
\bibitem{Galambos} J. Galambos, {\it The Asymptotic Theory of Extreme Order Statistics}, Malabar, FL: Krieger (1987)
\bibitem{BM} J.-P. Bouchaud and M. M\'ezard, {\it Universality classes for extreme-value statistics}, J. Phys. A: Math. Gen. {\bf 30}
7997 (1997)   
\bibitem{GBA} G. Ben Arous, L.V. Bogachev, S.A. Molchanov, {\it Limit theorems for sums of random exponentials}, Probability Theory and Related Fields,
{\bf 132} 579 (2005)
\bibitem{GBA2} G. Ben Arous, S.A. Molchanov, A.F. Ramirez, 
{\it Transition from the annealed to the quenched asymptotics for a random walk on random obstacles}, Ann. Probab. {\bf 33}, 
2149 (2005); {\it Phase transition asymptotics for random walks on a stationary random potential}, math.PR/0510519.
\bibitem{Derrida} B. Derrida, {\it Non-self averaging effects in sum of random variables}, in: On Three Levels, Ed M Fannes, C Maes
and A Verbeure (New York: Plenum, 1994) p 125.
\bibitem{CB} P. Cizeau and J.P. Bouchaud, {\it Theory of L\'evy Matrices}, Phys. Rev. {\bf E50} 1810 (1994) 
\bibitem{Burda} Z. Burda, J. Jurkiewicz, M. A. Nowak, G. Papp and I. Zahed, 
{\it Random L\'evy Matrices Revisited}, v3, cond-mat/0602087.
\bibitem{TW} C.A. Tracy and H. Widom, {\it Level spacing distributions and the Airy kernel}, Comm. Math. Phys., 
{\bf 159}, 33 (1994)
\bibitem{Spohn} for a review, see: H. Spohn, {\it Exact solutions for KPZ-type growth processes, random matrices, 
and equilibrium shapes of crystals}, cond-mat/0512011
\bibitem{Satya} S. Majumdar, S. Nechaev, {\it Exact asymptotic results for the Bernoulli matching model of sequence alignment}, Phys. Rev. E {\bf 72}, 020901(R) (2005) 
\bibitem{Johansson} K. Johansson. {\it Shape fluctuations and random matrices}, Comm. Math. Phys. {\bf 209} 437 (2000).
\bibitem{Sosh1} A. Soshnikov. {\it Universality at the edge of the spectrum in Wigner random matrices}, 
Comm. Math. Phys., {\bf 207} 697 (1999).
\bibitem{BBAP} J. Baik, G. Ben Arous, S. P{\'e}ch{\'e}, {\it Phase transition of the largest eigenvalue for non-null complex sample covariance matrices}, 
Ann. Probab. {\bf 33} 1643 (2005) 
\bibitem{Sosh2} A. Soshnikov. {\it Poisson statistics for the largest eigenvalues of Wigner 
random matrices with heavy tails}, Elect. Comm. in Probab. {\bf 9}, 82 (2004)
\bibitem{BBP} G. Biroli, J.-P. Bouchaud, Marc Potters, {\it On the top eigenvalue of heavy-tailed random matrices},
cond-mat/0609070, submitted to Europhysics Letters.
\bibitem{Prahofer} M. Pr\"ahofer,  H. Spohn, {\it Exact scaling functions for one-dimensional stationary KPZ growth},
J. Stat. Phys. {\bf 115} (1-2), 255-279 (2004).
\bibitem{MP} V.\ A.\ Mar\v{c}enko and L.\ A.\ Pastur, {\it Distribution of eigenvalues for some sets of
random matrices}, Math.\ USSR-Sb, {\bf 1}, 457-483 (1967)
\bibitem{Burda2} For a case where the {\it asymptotic} eigenvalue spectrum develops fat tails, see: 
Z. Burda, T. Gorlich, B. Waclaw, {\it Spectral properties of empirical covariance matrices for data with power-law tails}, physics/0603186  
\bibitem{Augusta} J.-P. Bouchaud, L. Laloux, M. A. Miceli, M. Potters, {\it Large dimension forecasting models and random singular value
spectra}, Eur. Phys. J. B (2006), b06036.
\bibitem{Zhang} Y. C. Zhang, {\it Growth anomaly and its implications}, Physica A{\bf 170}, 1 (1990)  
\bibitem{HH} T. Halpin Healey and Y.C. Zhang, {\it  Kinetic roughening, stochastic growth, directed polymers and all that}, 
Phys. Rep. {\bf 254} 189 (1995). 
\bibitem{BBLP} J.P. Bouchaud, E. Bouchaud, G. Lapasset, J. Plan\`es, {\it Models of fractal cracks}, Phys. Rev. Lett. {\bf 71}, 2240 (1993)
\bibitem{Hamley} B. Hambly, J. B. Martin, {\it Heavy tails in last-passage percolation}, math.PR/0604189.
\bibitem{DS} B. Derrida, H. Spohn, {\it Polymers on disordered trees, spin glasses, and traveling waves}, Journal of Statistical 
Physics, {\bf 51}, 817 (1988) 
\bibitem{PLDlog} for an interesting related discussion, see: D. Carpentier and P. Le Doussal, 
{\it Glass transition of a particle in a random potential, front selection in non linear renormalisation group}, Phys. Rev. E {\bf 63} 026110 (2001), 
and J-F Muzy, E.Bacry, A.Kozhemyak, {\it Extreme values and fat tails of multifractal fluctuations}, Phys. Rev. E {\bf 73} 066114 (2006).
\bibitem{Brock} D. Brockmann, L. Hufnagel, {\it Front Propagation in Reaction-Superdiffusion Dynamics}, cond-mat/0401322
\bibitem{Canet} for a recent discussion, see: L. Canet, M. A. Moore, {\it 
Mode-Coupling Theory of the Kardar-Parisi-Zhang Equation and the Functional Renormalization Group}, cond-mat/0604301
\bibitem{Kida} S. Kida, {\it Asymptotic properties of Burgers turbulence}, J. Fluid Mech. {\bf 93} 337 (1979)
\bibitem{PLD} P. Le Doussal, {\it Finite temperature FRG, droplets and decaying Burgers turbulence}, cond-mat/0605490, see also:
A. Middleton, P. Le Doussal, K. J. Wiese, {\it Measuring functional renormalization group fixed-point functions for pinned manifolds}, 
cond-mat/0606160
\bibitem{FH} D.S. Fisher and D.A. Huse, {\it Directed paths in a random potential}, Phys. Rev. B\textbf{\ 43}, 10728 (1991)
\bibitem{YS} M. Sales and H. Yoshino, {\it Fragility of the free-energy landscape of a directed polymer in random media}, Phys. Rev. B\textbf{\ 65}, 066131 (2002)
\bibitem{RdS} R. A. da Silveira and J.-P. Bouchaud, {\it Temperature and Disorder Chaos in Low Dimensional Directed Paths} 
Phys. Rev. Lett. {\bf 93}, 015901 (2004)
\bibitem{Deem} M. Deem, D. Chandler, {\it Classical diffusion in strong random media}, Journal of Statistical Physics, {\bf 76}, 911 (1994)
\bibitem{Dean1} D. S. Dean, I. Drummond, R. Horgan, {\it Perturbation schemes for flow in random media}, J. Phys. A: Math. Gen. {\bf 27}
5135 (1994) 
\bibitem{MB} see e.g. C. Monthus, J.-P. Bouchaud, {\it Trap Models and phenomenology of glasses}, J. Phys. A: Math. Gen. {\bf 29} 3847 (1996)
\bibitem{GBAC} G. Ben Arous, J. Cerny, {\it Dynamics of trap models}, math.PR/0603344 
\bibitem{Dean2} C. Touya, D. S. Dean, {\it Dynamical transition for a particle in a squared Gaussian potential}, cond-mat/0610470
\bibitem{MPV} M. M\'ezard, G. Parisi and M. A. Virasoro, {\it Spin Glass Theory and Beyond}, World Scientific, Singapore (1987).
\bibitem{Anderson} P. W. Anderson, in {\it Ill-condensed Matter}, Les Houches Lecture Notes (1978), World Scientific Pub.
\bibitem{Muller} see the very interesting recent paper: M. Mueller, S. Pankov, {\it Mean field theory for the 3D Coulomb glass}, cond-mat/0611021 
\bibitem{Parisi} G. Parisi, in {\it Chance \& Matter}, Les Houches Lecture Notes (1986), North-Holland 
\bibitem{ROM} G. Parisi, M. Potters, {\it Mean-Field Equations for Spin Models with Orthogonal Interaction Matrices}, 
J. Phys. A: Math. Gen. {\bf 28} 5267 (1995)
\bibitem{LSG} P. Cizeau, J.-P. Bouchaud, {\it Mean field theory of dilute spin-glasses with power-law interactions}, 
 J. Phys. A: Math. Gen. {\bf 26} L187 (1993)
\bibitem{Mertens} see e.g. H. Bauke, S. Franz, S. Mertens, {\it Number partitioning as a random energy model}, 
J. Stat. Mech. (2004) P04003.
\bibitem{ABM} A. Andreanov, F. Barbieri and O. C. Martin, {\it Large deviations in spin-glass ground-state energies}, EPJB {\bf 41},
365 (2004).
\bibitem{BKM} J.-P. Bouchaud, F. Krzakala, and O. C. Martin, {\it Energy exponents and corrections to scaling in Ising spin glasses},
Phys. Rev. B {\bf 68}, 224404 (2003)
\bibitem{Asp}  M. Goethe, T. Aspelmeier, {\it Free energy fluctuations in the mean-field Ising spin glass}, cond-mat/0610228  and refs. therein.
\bibitem{Campbell} L. W. Bernardi and I. A. Campbell, {\it Critical exponents in Ising spin glasses},  Phys. Rev. B {\bf 56} 5271 (1997)
\bibitem{Rammal} R. Rammal, {\it Spin dynamics and glassy relaxation on fractals and percolation structures}, Journal de physique, {\bf 46}, 1835 (1985)
\bibitem{Vinokur} V. Vinokur, C. Marchetti, L.W. Chen, {\it Glassy Motion of Elastic Manifolds}, Phys. Rev. Lett. {\bf 77}, 1845(1996)

\end{thebibliography}
\end{document}